\documentclass[10pt,a4paper]{article}

\usepackage[utf8]{inputenc}
\usepackage[T1]{fontenc}
\usepackage{aeguill}
\usepackage{lmodern}
\usepackage[english]{babel}
\AtBeginDocument{\selectlanguage{english}}
\usepackage{csquotes}
{\rmfamily
  \DeclareFontShape{T1}{lmr}{b}{sc}{<->ssub*cmr/bx/sc}{}%
  \DeclareFontShape{T1}{lmr}{bx}{sc}{<->ssub*cmr/bx/sc}{}}
\usepackage{amsmath}
\usepackage{amssymb}
\usepackage{latexsym}
\usepackage{stmaryrd} 
\usepackage[nointegrals]{wasysym} 
\usepackage{dsfont}
\usepackage{graphicx}
\usepackage{pdflscape} 
\usepackage{calc}
\usepackage{booktabs}
\usepackage{chemarrow}
\usepackage{mathtools} 
\usepackage{wasysym}
\usepackage{sidecap}
\usepackage{array}
\usepackage{boxedminipage}
\usepackage{setspace} 

\usepackage{footnotehyper}
\newenvironment{landscapetable}{\begin{landscape}\begin{table}}
    {\end{table}\end{landscape}}
\makesavenoteenv{landscapetable}
\usepackage{fancyhdr}
\usepackage{comment}
\usepackage{color,soul} \sethlcolor{yellow}
\usepackage{enumerate}
\usepackage{multirow}
\usepackage{algorithmic}
\usepackage{xcolor}
\usepackage{float}
\usepackage{alltt}
\usepackage{upquote}
\usepackage{listings}
\definecolor{dkgreen}{rgb}{0,0.4,0}

\newcommand{\YAMLcolonstyle}{\color{red}}
\newcommand{\YAMLkeystyle}{\color{purple}}
\newcommand{\YAMLvaluestyle}{\color{black}}
\newcommand{\YAMLstringstyle}{\color{dkgreen}}
\newcommand{\YAMLcommentstyle}{\color{cyan}}
\newcommand{\YAMLkeywordstyle}{\bfseries\color{violet}}

\makeatletter
\newcommand*\language@yaml{yaml}

\expandafter\expandafter\expandafter\lstdefinelanguage
\expandafter{\language@yaml}
{
  keywords={true,false,yes,no,null}, 
  keywordstyle=\YAMLkeywordstyle,
  basicstyle=\ttfamily\YAMLkeystyle, 
  breaklines=false,
  columns=[l]fullflexible,
  keepspaces=true,
  sensitive=false,
  morecomment=[l]{\#},
  commentstyle=\YAMLcommentstyle,
  stringstyle=\YAMLstringstyle,
  moredelim=[l][\color{orange}]{\&},
  moredelim=[l][\color{magenta}]{*},
  moredelim=**[il][\YAMLcolonstyle{:}\YAMLvaluestyle]{:},
  morestring=[b]',
  morestring=[b]",
  literate =    {---}{{\ProcessThreeDashes}}3
                {>}{{\YAMLcolonstyle\textgreater}}1
                {|}{{\YAMLcolonstyle\textbar}}1
                {\ \ -\ }{{\ \ {\YAMLcolonstyle-}\ }}3,
}
\makeatother

\newcommand\ProcessThreeDashes{\mbox{\color{cyan}\ttfamily-{-}-}}

\lstdefinelanguage{docker}{
  keywords={FROM, RUN, COPY, ADD, ENTRYPOINT, CMD,  ENV, WORKDIR, EXPOSE, LABEL, USER, VOLUME, STOPSIGNAL, ONBUILD, MAINTAINER},
  keywordstyle=\color{purple}\bfseries,
  identifierstyle=\color{black},
  sensitive=false,
  comment=[l]{\#},
  commentstyle=\color{cyan}\ttfamily,
  stringstyle=\color{dkgreen}\ttfamily,
  morestring=[b]',
  morestring=[b]"
}

\lstdefinelanguage{docker-compose}{
  keywords={image, environment, ports, container_name, ports, volumes, links},
  keywordstyle=\color{purple}\bfseries,
  identifierstyle=\color{black},
  sensitive=false,
  comment=[l]{\#},
  commentstyle=\color{cyan}\ttfamily,
  stringstyle=\color{dkgreen}\ttfamily,
  morestring=[b]',
  morestring=[b]"
}
\lstdefinelanguage{docker-compose-2}{
  keywords={version, volumes, services},
  keywordstyle=\color{dkgreen}\bfseries,
  keywords=[2]{image, environment, ports, container_name, ports, links, build},
  keywordstyle=[2]\color{purple}\bfseries,
  identifierstyle=\color{black},
  sensitive=false,
  comment=[l]{\#},
  commentstyle=\color{cyan}\ttfamily,
  stringstyle=\color{dkgreen}\ttfamily,
  morestring=[b]',
  morestring=[b]"
}

\lstnewenvironment{dockerfile}[1][]%
{\smallbreak\noindent\ignorespaces\lstset{language={docker},frame=none,#1}}{}%
\lstnewenvironment{dockerfilec}[1][]%
{\smallbreak\noindent\ignorespaces\lstset{language={docker},frame=single,#1}}{}%

\usepackage{tikz}
\usetikzlibrary{tikzmark}
\usetikzlibrary{decorations.pathreplacing,calligraphy}
\usetikzlibrary{patterns}
\usetikzlibrary{matrix,arrows,decorations.pathmorphing}
\usetikzlibrary{shapes.geometric,shapes.misc}
\usetikzlibrary{shadows}
\usetikzlibrary{backgrounds}
\usetikzlibrary{shapes.callouts}
\usetikzlibrary{decorations.text}
\usepackage{smartdiagram}
\usesmartdiagramlibrary{additions}

\usepackage{hyperref}
\hypersetup{%
  pdftitle={Docker-based CI/CD for Rocq/OCaml projects},%
  pdfauthor={Erik Martin-Dorel},%
  pdfsubject={},%
  pdfkeywords={},%
  pdfstartview=Fit,
  pdfpagelayout=OneColumn,
  pdfpagemode=UseOutlines,
  colorlinks=false,urlcolor=blue}

\usepackage{xurl} 
\usepackage{makeidx}
\usepackage[open=true,openlevel=0,numbered=true]{bookmark}

\usepackage{pmboxdraw}

\let\origdoublepage\cleardoublepage
\newcommand{\clearemptydoublepage}{%
  \clearpage
  {\pagestyle{empty}\origdoublepage}%
}
\let\cleardoublepage=\clearemptydoublepage
\makeatletter
\newcommand{\clearevenpage}{\clearpage
  {\pagestyle{empty}%
    \ifodd\c@page
    \hbox{}\newpage\if@twocolumn\hbox{}\newpage\fi\fi
  }%
}
\makeatother

\newcommand{\YML}[1]{\lstinline[language=yaml]`#1`}
\newcommand{\C}[1]{\mbox{\textup{\lstinline`#1`}}} 
\newcommand{\G}[1]{\lstinline`#1`}

\newcounter{regbmctr}
\newcommand{\registerbm}[1]{%
  \phantomsection
  \stepcounter{regbmctr}%
  \hypertarget{regbm:\theregbmctr}{}%
  \bookmark[level=1,dest={regbm:\theregbmctr}]{#1}%
  \markboth{\MakeUppercase{#1}}{\MakeUppercase{#1}}%
}
\newcommand{\registerbmtoc}[1]{%
  \phantomsection
  \addcontentsline{toc}{section}{#1}%
  \markboth{\MakeUppercase{#1}}{\MakeUppercase{#1}}%
}

\providecommand*{\email}[1]{\href{mailto:#1}{\nolinkurl{#1}}}
\providecommand{\bsc}{\textsc}
\newcommand*{\OPAM}{\textsc{opam}}
\newcommand*{\LOC}{\textsc{loc}}

\usepackage{tocloft}

\author{\'Erik \bsc{Martin-Dorel}\\\email{erik@martin-dorel.org}\\{Independent Researcher, Toulouse, France}}
\title{Docker-based CI/CD for Rocq/OCaml projects}
\date{20 October 2025}
\begin{document}

\maketitle

\begin{abstract}
  This paper presents three closely-related software projects, namely:
  \{docker-coq, docker-coq-action, docker-keeper\}. It aims at two
  objectives: provide a high-level description of the available
  features---to foster the use of a Docker-based CI/CD for Rocq
  (formerly known as Coq) or OCaml projects---and document the
  underlying requirements and the main design choices of these three
  DevOps tools---to help their future maintainers.
\end{abstract}

\registerbm{\contentsname}
\tableofcontents

\clearpage

\section{Introduction}
This paper has been written in 2024 (while the author was with
lab.~\textsc{irit} at \emph{Université de Toulouse}) then revised in
October 2025. It presents three closely-related software projects,
namely docker-coq, docker-coq-action, and docker-keeper. It aims at
two objectives: provide a high-level description of the available
features---to foster the use of a Docker-based CI/CD for Rocq or OCaml
projects---and document the underlying requirements and the main
design choices of these DevOps tools---to help the future docker-rocq
maintainers.

Thanks to his teaching activities in the Software Engineering Master
of Univ.~Toulouse (\emph{Master SDL}), the author have become fluent
with the \emph{CI/CD} approach as well as with \emph{OS-level
  virtualization using Docker}, which he~then applied in his research
in the Rocq and OCaml communtities.

In the sequel, we will recap the main features of these two
methodologies in Section~\ref{sec:docker-bg}, then present our
software engineering contributions in this scope: docker-coq in
Section~\ref{sec:docker-coq}, docker-hub-helper in
Section~\ref{sec:docker-hub-helper}, docker-keeper in
Sections~\ref{sec:docker-keeper-overview}
and~\ref{sec:docker-keeper-inner}, docker-coq-action in
Sections~\ref{sec:docker-coq-action-overview}
and~\ref{sec:docker-coq-action-inner}, while
Section~\ref{sec:docker-coq-comparison} will be devoted to a
qualitative evaluation of this Docker-based approach, comparing with
other state-of-the-art approaches. Then, Section~\ref{sec:conclusion}
provides conclusions and perspectives. The work described in this
paper has been first presented in the Coq Workshop
2020~\cite{Coq2020}.

In March 2025, the \emph{Coq proof assistant} has been renamed as the
\emph{Rocq
  prover}\footnote{cf.~\url{https://rocq-prover.org/about\#Name}}
(version 9.0.0) but for the sake of simplicity, we will use the
initial names throughout the paper, namely (Coq \& docker-coq \&
docker-coq-action) instead of (Rocq \& docker-rocq \&
docker-opam-action).

\section{Background (CI/CD \& Docker)}
\label{sec:docker-bg}

\begin{figure}[!ht]
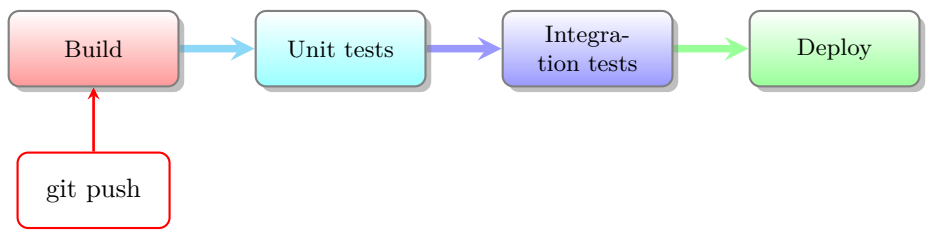

  \smartdiagramset{back arrow disabled=true, text width=2cm, module x sep=3.25,
    additions={additional item offset=0.85cm,additional item border color=red,additional connections disabled=false,additional arrow color=red,additional arrow tip=stealth,additional arrow line width=1pt
    }}
  \smartdiagramadd[flow diagram:horizontal]{
    Build, {Unit tests}, {Integration tests}, Deploy}%
  {below of module1/{git push}}
  \bigbreak
  \bigbreak
  \bigbreak
  \bigbreak
  \caption[Example of a 4-stage CI/CD pipeline]{Example of a 4-stage CI/CD pipeline.}
  \label{fig:CI-pipeline}
\end{figure}

The term ``CI'' has been coined by
Grady~Booch~\cite[p. 256]{booch1994object} in the context of software
engineering with object-oriented programming languages, which then
became a key practice of extreme programming~(cf.~Beck~\cite{Beck1999}) then of
agile methodologies in general, aiming to ``deliver working software
frequently''.  Martin Fowler provides a very detailed description of
CI, divided in 11 practices (see Figure~\ref{fig:CI}).
\begin{figure}[!ht]
  \begin{small}
\fbox{\begin{minipage}{\textwidth}\begin{enumerate}
\item maintain a single source repository using version control;
\item ``automate the build'' using dedicated tools;
\item include automated tests in the build process;
\item integrate developed code in the main branch frequently;
\item ensure the main branch always builds, using a dedicated machine;
\item ``fix broken builds immediately'';
\item ``keep the build fast'' (if need be, using so-called pipelines);
\item ``hide work-in-progress'' that is not production-ready (using feature flags or so);
\item ``test in a clone of the production environment'';
\item ensure the build status is clearly visible;
\item ``automate deployment'' into production.
\end{enumerate}\end{minipage}}
\caption[Practices of Continuous Integration]{Practices of Continuous Integration~(cf.~Fowler~\cite{Fowler:CI}).}
\label{fig:CI}
\end{small}
\end{figure}
Briefly speaking, CI is a general methodology that aims at
\mbox{systematically} and automatically running jobs for each push in the
main development branch. These jobs are generally run in a
continuous integration platform (e.g., Jenkins, Travis CI, CircleCI,
GitLab CI, GitHub Actions, etc.),
and organized in pipelines containing several
stages, typically for building the project, running unit tests,
integration tests, and so on (see Figure~\ref{fig:CI-pipeline}).
A job failure prevents running the subsequent stages, and sets an
error status for the pipeline. CI makes it possible to detect
integration issues or regressions very early in the development, and
thereby reduces the cost of software bugs (incidentally, CI also
significantly improves developer experience). Thus to some extend, CI
can be viewed as a V~\&~V technique that should be continuously
applied in the software lifecycle, namely for each commit of the main
development branch, and typically also for feature branches within a
``pull request'' process, in order to provide pull-request reviewers
with more insight about the correctness of the candidate
contribution. The last step that can occur in a CI pipeline is named
``CD'' (so ``CI/CD'' is often used to denote the whole practice), but
it should be noted that CD can denote two slightly different
approaches: $(i)$ Continous Delivery, where the deployment to
production is only done on-demand, after validation steps such as
manual review or acceptance tests in a pre-production environment;
$(ii)$ Continuous Deployment, where a successful CI pipeline
automatically triggers a deployment to production (so that in this
latter approach, the confidence in the testing suite is paramount).

In the context of Coq projects, even if a successful build means
already ``more'' than a mere compilation in a conventional programming
language, due to the presence of accompanying Coq proofs, a CI
pipeline can involve all these standard stages, namely,
\emph{building} the project with \texttt{coqc}, running end-to-end
\emph{tests}, benchmarks, or compatibility checks,\footnote{A good
  example is given by the CI infrastructure of Coq itself, where a
  systematic testing of reverse dependencies has been applied with
  Théo~Zimmermann's work~\cite[Chap. 3.6]{Zimmermann:PhD}.} then
automatically \emph{deploying} artifacts, which can just as well be
documentation artifacts.
For any Coq project that follows one such CI/CD approach, a
valuable insight provided by the CI amounts to telling which tested
versions of Coq are compatible. (For the record, the current release
cycle of Coq yields a major release about every six
months.\footnote{cf.~\url{https://github.com/coq/coq/blob/master/CONTRIBUTING.md\#beta-testing}})

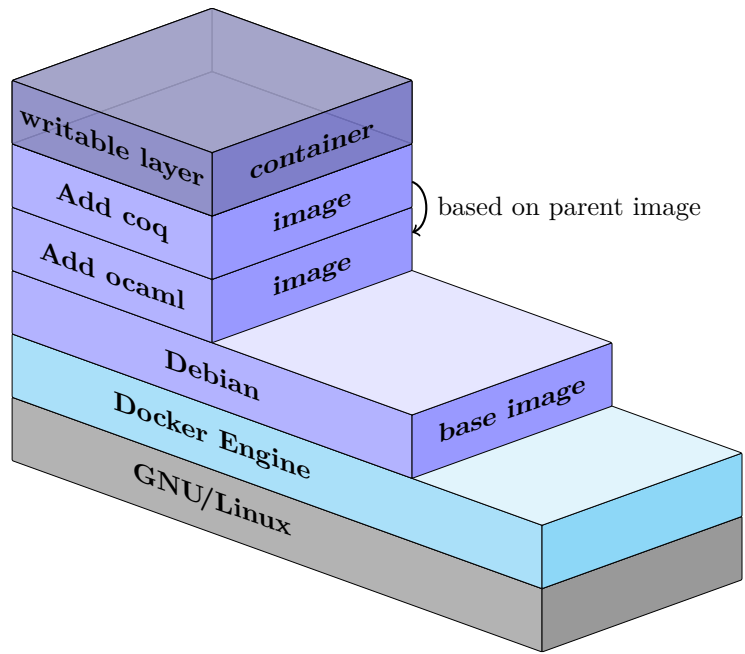
\begin{figure}[!t]
\def\xsl{0.005}
\def\ysl{0.36}
\let\siz=\normalsize
\newcommand{\elemcube}[6][blue]{	
        \draw [fill=#1!30,very thin,rounded corners=0.5pt] (#2+1,#3,#4) -- ++(0,1,0) -- ++(0,0,-1) -- ++(0, -1, 0) -- cycle;
        \draw [fill=#1!40,very thin,rounded corners=0.5pt] (#2,#3+1,#4) -- ++(1,0,0) -- ++(0,0,-1) -- ++(-1, 0, 0) -- cycle; 
        \draw [fill=#1!10,very thin,rounded corners=0.5pt] (#2,#3,#4)   -- ++(1,0,0) -- ++(0,1,0)  -- ++(-1, 0, 0) -- cycle;
        \node [xslant=-\xsl, yslant=-\ysl, font={\siz\bfseries}]at (#2+1, #3+0.5, #4-0.5) {#5};
        \node [xslant=\xsl, yslant=\ysl, font={\siz\bfseries}]at (#2+0.5, #3+1, #4-0.5) {#6};
}
\newcommand{\elemcubetwo}[6][blue]{	
        \draw [fill=#1!30,very thin,rounded corners=0.5pt] (#2+1,#3,#4) -- ++(0,2,0) -- ++(0,0,-1) -- ++(0, -2, 0) -- cycle;
        \draw [fill=#1!40,very thin,rounded corners=0.5pt] (#2,#3+2,#4) -- ++(1,0,0) -- ++(0,0,-1) -- ++(-1, 0, 0) -- cycle; 
        \draw [fill=#1!10,very thin,rounded corners=0.5pt] (#2,#3,#4)   -- ++(1,0,0) -- ++(0,2,0)  -- ++(-1, 0, 0) -- cycle;
        \node [xslant=-\xsl, yslant=-\ysl, font={\siz\bfseries}]at (#2+1, #3+1, #4-0.5) {#5};
        \node [xslant=\xsl, yslant=\ysl, font={\siz\bfseries}]at (#2+0.5, #3+2, #4-0.5) {#6};
}
\newcommand{\elemcubetwoh}[6][blue]{	
        \draw [fill=#1!30,very thin,rounded corners=0.5pt] (#2+1,#3,#4) -- ++(0,2.65,0) -- ++(0,0,-1) -- ++(0, -2.65, 0) -- cycle;
        \draw [fill=#1!40,very thin,rounded corners=0.5pt] (#2,#3+2.65,#4) -- ++(1,0,0) -- ++(0,0,-1) -- ++(-1, 0, 0) -- cycle; 
        \draw [fill=#1!10,very thin,rounded corners=0.5pt] (#2,#3,#4)   -- ++(1,0,0) -- ++(0,2.65,0)  -- ++(-1, 0, 0) -- cycle;
        \node [xslant=-\xsl, yslant=-\ysl, font={\siz\bfseries}]at (#2+1, #3+1, #4-0.5) {#5};
        \node [xslant=\xsl, yslant=\ysl, font={\siz\bfseries}]at (#2+0.5, #3+2.65, #4-0.5) {#6};
}
\newcommand{\elemcubetransp}[6][blue]{	
	\filldraw[fill opacity=0.7,rounded corners=0.5pt] (#2,#3,#4-1) -- ++(1,0,0) -- ++(0,1,0) -- ++(-1, 0, 0) -- cycle;
	\filldraw[fill opacity=0.7,rounded corners=0.5pt] (#2,#3,#4-1) -- ++(1,0,0) -- ++(0,0,1) -- ++(-1, 0, 0) -- cycle;
	\filldraw[fill opacity=0.7,rounded corners=0.5pt] (#2,#3,#4-1) -- ++(0,1,0) -- ++(0,0,1) -- ++(0, -1, 0) -- cycle;
	\filldraw[fill opacity=0.7,fill=#1!20,rounded corners=0.5pt] (#2+1,#3+1,#4) -- ++(-1,0,0) -- ++(0,-1,0) -- ++(1, 0, 0) -- cycle;
	\filldraw[fill opacity=0.7,fill=#1!40,rounded corners=0.5pt] (#2+1,#3+1,#4) -- ++(-1,0,0) -- ++(0,0,-1) -- ++(1, 0, 0) -- cycle;
	\filldraw[fill opacity=0.7,fill=#1!30,rounded corners=0.5pt] (#2+1,#3+1,#4) -- ++(0,-1,0) -- ++(0,0,-1) -- ++(0, 1, 0) -- cycle;
        \node [xslant=-\xsl, yslant=-\ysl, font={\siz\bfseries}]at (#2+1, #3+0.5, #4-0.5) {#5};
        \node [xslant=\xsl, yslant=\ysl, font={\siz\bfseries}]at (#2+0.5, #3+1, #4-0.5) {#6};
}
\newenvironment{stackedboxes}{
  \begin{center}%
  \begin{tikzpicture}[x=(200:3.5cm), y=(-20:3.5cm), z=(90:1.05cm),scale=0.8]%
  }{\end{tikzpicture}%
  \end{center}}
\newcommand{\drawbasearrow}[2]{%
  \draw[->,thick] (0,1,#1+0.4) to[bend right=-70] (0,1,#1-0.4);
  \draw node[right] at (-0.03,1.05,#1) {#2};
}
\begin{stackedboxes}
	\elemcubetwoh[black]{0}{0}{-1}{GNU\!/Linux}{}
	\elemcubetwoh[cyan]{0}{0}{0}{Docker Engine}{}
	\elemcubetwo{0}{0}{1}{Debian}{base image}
	\elemcube{0}{0}{2}{Add ocaml}{image}
	\elemcube{0}{0}{3}{Add coq}{image}
	\elemcubetransp{0}{0}{4}{writable layer}{container}
        \drawbasearrow{2}{based on parent image}
\end{stackedboxes}
\caption[Docker's filesystem architecture]{Docker's filesystem architecture: it relies on layered images
  (which are read-only), a common layer can be shared between several
  images, and will be stored only once; and there are containers
  (starting from an image, endowed with a writable layer), which can
  in turn be ``snapshotted'' to get new images.}
  \label{fig:docker-layers}
\end{figure}

\begin{figure}[!ht]
  \includegraphics[width=\textwidth]{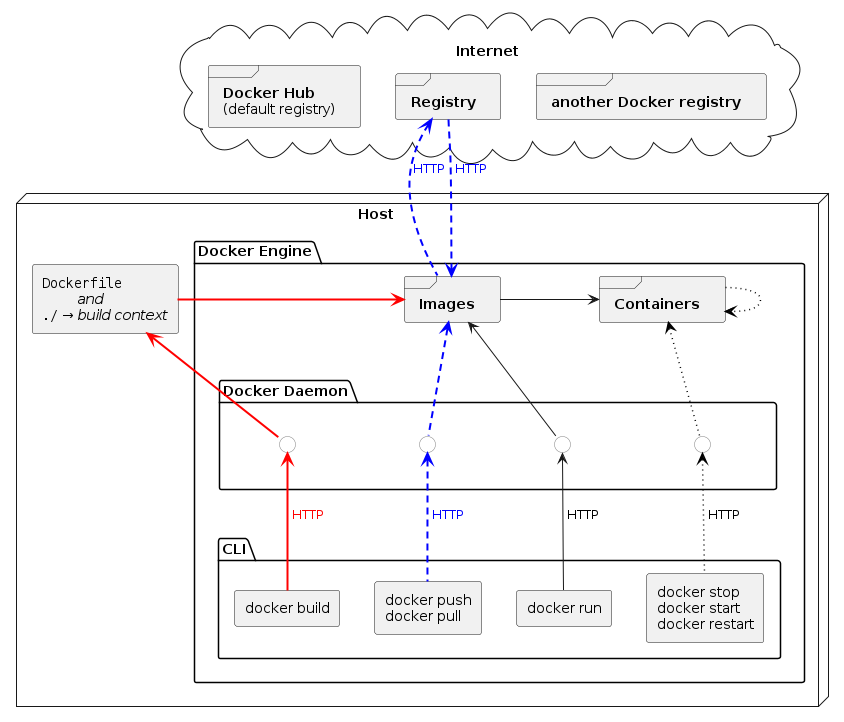}
  \caption[Docker Engine's architecture]{Docker Engine's architecture: four major commands are
    exemplified, with a different arrow style for each of them; images
    can be built on the host (using a \texttt{Dockerfile} along with the
    so-called \emph{build context}), pushed or pulled from a remote
    \emph{Docker registry}, and run locally to get a container that
    can then be stopped, restarted, and so on.}
\label{fig:docker-engine}
\end{figure}

As mentioned earlier in this section, several continuous integration
platforms are available: roughly speaking, they consist in a dedicated
server (that can be on-premise or hosted by a cloud provider) that is
in charge of running pipelines (fetching the relevant code, building
it, and running tests), ideally in an isolated way. This isolation can
be achieved by virtualization techniques, either using hypervisors
(VMs) or OS-level virtualization (Docker\footnote{First released in
  March 2013 as free software.} containers).  Generally speaking, the
use of containers has several advantages over the use of mere VMs:
containers are lightweight in terms of disk, memory, and CPU usage,
given they do not emulate a whole OS, but just consist in Linux
processes that are isolated by relying on specific features of the
Linux kernel (in particular: isolation w.r.t.\ other processes and
network, thanks to namespaces; limiting resources usage
(memory/CPU/etc.), thanks to control groups; limiting root privileges
at a very granular level, thanks to capabilities; isolation w.r.t.\
the host filesystem, thanks to ``chroot jails''). Beyond the notion of
container and the corresponding isolation features that Docker makes
easy to use, Docker introduces the notion of image (roughly, a
snapshot of a container's filesystem).
Figure~\ref{fig:docker-layers}
gives an overview of Docker's filesystem architecture on the local
host, while Figure~\ref{fig:docker-engine} summarizes the overall
architecture of Docker Engine w.r.t.\ remote Docker registries, along
with the impact of the main command-line requests on the lifecycle of Docker containers.

Finally, it can be noted that using Docker containers in the context of
CI makes a lot of sense, as they make it possible to enhance
reproducibility, or performance in a portable manner. To be more
precise, a Docker-based CI/CD pipeline can successively build a Docker
image, test it, then deploy it in production, with the guarantee that
the deployed artifact is exactly the same as the one that was tested
in the pipeline (cf.~Practice 8 of Figure~\ref{fig:CI}). Besides, another
typical approach (which will be instrumental in the upcoming
Section~\ref{sec:docker-coq}) is the ability to prepare dedicated
Docker images where the main project's dependencies are prebuilt. Then,
the CI pipelines are run starting from these images. This Docker-based
approach is compatible with most continuous integration platforms
(whatever is their underlying OS, thanks to Docker's portability,
e.g., we can build a Debian-based image on an Ubuntu host, then run it
on a Fedora or a Windows 11 host), and above all, this makes it
possible to significantly reduce the overall CI time.

\section{docker-base, docker-coq, and docker-mathcomp}
\label{sec:docker-coq}

Given Coq is mostly written in OCaml which comes with its own
source-based package manager
(\OPAM\footnote{cf.~\url{https://opam.ocaml.org/}}), a natural way to
install Coq itself, as well as Coq projects, consists in using
\OPAM. We can then rely on Docker to get binary packages of Coq in a
given Linux distribution (say, Debian stable) and distribute them on
Docker Hub. This is precisely the aim of the docker-coq
project~\cite{docker-coq}. And likewise for the docker-mathcomp
project~\cite{docker-mathcomp} (a prebuilt \OPAM{} switch with \texttt{coq} and
\texttt{coq-mathcomp-character} packages). Initially, their development started
with the need to build the ValidSDP project~\cite{validsdp} within
Travis~CI (the \emph{de facto} standard CI platform for GitHub in that
time), which imposed a 50' time limit for each build. Given that
ValidSDP depends on a large number of dependencies (Coq + nine OCaml
or Coq libraries), it was paramount to be able to prebuild a large
part of the dependencies (here, Coq and MathComp). Although Travis~CI
did not directly support Docker,%
\footnote{A Travis~CI issue is still open on this very topic:
  \url{https://github.com/travis-ci/travis-ci/issues/7726}} but
allowed to spin a Docker service and manually call %
\texttt{docker run} or related commands, the Docker-based approach
looked promising (while Travis' native caching features were
useless, because for a CI caching to be successful, the build needs to
proceed at least once).  For the record, two different templates for
Travis~CI were ultimately written.%
\footnote{cf.~\url{https://github.com/erikmd/docker-coq-travis-ci-demo-1/blob/master/.travis.yml}
  \&
  \url{https://github.com/erikmd/docker-coq-travis-ci-demo-2/blob/master/.travis.yml}}
They were functional albeit quite ``verbose'' (especially the first
template ``docker-coq-travis-ci-demo-1'', because of the explicit
presence of docker arguments as well as the various quoting levels
involved).

\begin{SCfigure}[2.0][!ht]
    \fbox{\includegraphics[width=0.4\textwidth]{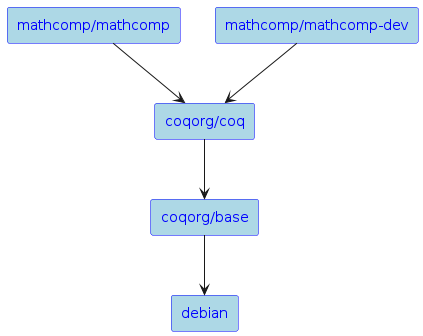}}
    \caption[Hierarchy of the (docker-mathcomp) images]{Hierarchy of the images: arrows ``$x \chemarrow y$''
      stand for
      ``$x$ has as parent image $y$''. \linebreak Regarding Docker's
      syntax of image names: they have the form
      \texttt{repo\string:tag} where \texttt{repo} is either
      ``\texttt{some-domain.tld/user/name}'' %
      ($2$~slashes or more) for custom registries,
      ``\texttt{user/name}'' ($1$ slash) for the default registry $=$
      Docker Hub, or ``\texttt{name}'' (no slash) for official images,
      such as \texttt{ubuntu}, \texttt{debian}, \texttt{python},
      \texttt{mysql}, etc.}
    \label{fig:docker-mathcomp}
\end{SCfigure}

As a prerequisite of this Travis CI based setup, my first experiments
with Docker Hub to ``dockerize'' Coq started in August 2018; I
contacted the Coq team
on\footnote{cf.~\url{https://github.com/coq/coq/issues/8474}}
September 13, 2018 to suggest one distributes images of Coq for both
stable and \texttt{dev} versions, then a fruitful discussion was
launched. We announced the availability of the \texttt{coqorg/coq}
images on the coq-club mailing list
in\footnote{cf.~\url{https://sympa.inria.fr/sympa/arc/coq-club/2018-11/msg00035.html}}
November 2018.
Since then, I have been maintaining four different images
repositories, which are summarized in Figure~\ref{fig:docker-mathcomp}
and Table~\ref{tab:docker-mathcomp}.

\vspace{-.5em}
\begin{table}[!ht]
  \begin{footnotesize}
  \begin{tabular}{@{}rlll@{}} \toprule
    Docker Hub repository & build infrastructure & GitHub repository with Dockerfiles \\ \midrule
    \href{https://hub.docker.com/r/mathcomp/mathcomp-dev}{\nolinkurl{mathcomp/mathcomp-dev}}
    & custom CI/CD
    & \href{https://github.com/math-comp/math-comp}{\nolinkurl{github.com/math-comp/math-comp}} \\
    \href{https://hub.docker.com/r/mathcomp/mathcomp}{\nolinkurl{mathcomp/mathcomp}}
    & docker-keeper
    & \href{https://github.com/math-comp/docker-mathcomp}{\nolinkurl{github.com/math-comp/docker-mathcomp}} \\
    \href{https://hub.docker.com/r/coqorg/coq}{\nolinkurl{coqorg/coq}}
    & docker-keeper
    & \href{https://github.com/coq-community/docker-coq}{\nolinkurl{github.com/coq-community/docker-coq}} \\
    \href{https://hub.docker.com/r/coqorg/base}{\nolinkurl{coqorg/base}}
    & docker-keeper
    & \href{https://github.com/coq-community/docker-base}{\nolinkurl{github.com/coq-community/docker-base}} \\
    \href{https://hub.docker.com/\_/debian}{\nolinkurl{debian}}
    & official images & omitted for brevity \\
    \bottomrule
  \end{tabular}
  \end{footnotesize}
\caption[URL of the Docker Hub repositories at stake]{URL of the Docker image repositories and their respective
  Dockerfile repositories. The current infrastructure strongly relies
  on the docker-keeper tool, which will be presented in
  Section~\ref{sec:docker-keeper-overview}.}
\label{tab:docker-mathcomp}
\end{table}

\section{docker-hub-helper}
\label{sec:docker-hub-helper}

Contrarily to source tarballs associated to each release of Coq, the
Docker images built from a given release are intended to be rebuilt on
a regular basis, given $(i)$ they incorporate a whole Linux
distribution, which thereby benefits from package updates
(including--but not limited to--security updates) and $(ii)$ they also
need to integrate new versions of some essential tools and libraries
(including the \OPAM{} binary, the OCaml compiler, or the
\texttt{dune} package). Hence the need for some tooling to automate
part of the process.

The first approach (that was applied from November 2018 until June
2020) relied on Docker Hub's standard feature called ``automated
builds''. This feature consists in associating the Docker Hub
repository with a GitHub repository, and reading the Dockerfile
(containing the precise Coq version and all build details) from a
dedicated branch. In the sequel, we will refer to this repository
model as the \emph{multi-branches infrastructure}.

For all Coq versions, given the Dockerfiles were almost identical but
not fully, a ${\sim}500$-\LOC{} Python~3 helper tool
(docker-hub-helper~\cite{docker-hub-helper}) was written to help
maintaining both Coq and MathComp Docker Hub repositories, handling
Git branches semi-automatically as well as HTTP requests to trigger
rebuilds.

\section{docker-keeper: objectives and interface}
\label{sec:docker-keeper-overview}

The approach mentioned in the previous section was unsatisfactory, due
to two main concerns. First, it did not scale, regarding performance
(e.g., building docker-mathcomp 1.11.0 for 6 versions of
Coq was taking more than 7 hours) and usability (despite the
availability of the docker-hub-helper CLI tool, many clicks in the Docker Hub web UI were needed).
Second, the Git setup for this multi-branches infrastructure
was incompatible with a pull requests workflow (a global change
to the Dockerfile had to be ported to many different Git branches).

Hence the need to a new infrastructure and tooling. On the one hand, using a CI/CD platform such as GitLab CI looked appealing given the large amount of CI minutes offered in the free tier for public projects (although this amount has been significantly reduced since then, it is still tractable for docker-coq to date).

On the other hand, it appeared necessary to adopt a single-branch architecture for the main Git repository that gathers the configuration of the Docker images to build.
We chose to store these configurations in a single YAML file, endowed
with a domain-specific language that encodes the various features
required for maintaining docker-coq. This YAML specification is then
parsed at CI runtime by a dedicated tool, that we chose to name
docker-keeper~\cite{docker-keeper}. The overall project amounts to
${\sim}2.2k$\LOC{} of Python-3/Bash/GitLab-CI code, including tests.
When presenting this work at the Coq Workshop 2020~\cite{Coq2020},
we identified the following non-functional requirements that this
approach addresses:
\begin{description}
\item[scalability] regarding release build time (but also regarding global rebuilds time, given the number of supported versions typically increases over time) 
\item[maintainability] by operating on a single YAML file stored in the main branch
\item[genericity] as it's worth it to %
  be applicable to non-Coq projects as well
\item[consistency] of the deployed images w.r.t.\ the YAML spec (achieved by detecting obsolete tags to be removed from Docker Hub, and by checking injectivity of the image-to-tag mapping, so different images cannot be mapped to duplicate tags)
\item[transparency] regarding the CI logs (they were not public in Docker Hub's automated builds, unlike GitLab CI logs)
\item[better documentation] regarding the metadata of each image, and
  providing an automatically-generated list of available
  tags.\footnote{cf.~\url{https://hub.docker.com/r/coqorg/coq\#supported-tags}}
\end{description}

The sequel of this section aims to provide an overall presentation of the use
of the docker-keeper tool, while section~\ref{sec:docker-keeper-inner}
will elaborate on its inner workings.

Figure~\ref{fig:docker-keeper-template-tree} shows the
contents of a \texttt{docker-keeper}-based repository (e.g., obtained
by cloning the \texttt{docker-keeper-template}
repository~\cite{docker-keeper-template}). A key element is
the GitLab CI config file (\texttt{.gitlab-ci.yml},
cf.~Figure~\ref{fig:docker-keeper-template-ci-yml}), which just
amounts to including the generic config stored in the
\texttt{external/docker-keeper} subdirectory, which itself reads the
3 main files having tag ``{\color{cyan}\small\textbf{\texttt{TO BE CUSTOMIZED}}}'' in
Figure~\ref{fig:docker-keeper-template-tree}.

\begin{figure}[!htb]
  \newcommand*{\nextFigur}{\ref{fig:docker-keeper-template-ci-yml}}
  \small
\!\!\!\begin{boxedminipage}{1.05\textwidth}
\begin{alltt}
fork of docker-keeper-template
├── external/docker-keeper/ {\color{cyan}# Git subtree (omitted here)}
├── images.yml     {\color{cyan}# main YAML file = docker-keeper DSL \textbf{(TO BE CUSTOMIZED)}}
├── LICENSE        {\color{cyan}# MIT license by default}
├── README.md      {\color{cyan}# main documentation file & template \textbf{(TO BE CUSTOMIZED)}}
├── stable
│   └── Dockerfile {\color{cyan}# main config to build Docker images \textbf{(TO BE CUSTOMIZED)}}
├── .dockerignore  {\color{cyan}# needed by Docker best practices; contains * by default}
├── .gitignore     {\color{cyan}# needed by Git best practices}
├── .gitlab-ci.yml {\color{cyan}# default GitLab CI config; see Figure} \nextFigur{}
└── .git/          {\color{cyan}# Git repository}
\end{alltt}
\end{boxedminipage}
\caption[Overview of the main files of docker-keeper-template]{Overview of the main files of the
  \texttt{docker-keeper-template} Git repository~%
  \cite{docker-keeper-template}.}
\label{fig:docker-keeper-template-tree}
\end{figure}
\begin{figure}[!htb]
  \small
\!\!\!\begin{boxedminipage}{1.05\textwidth}
\begin{lstlisting}[language=yaml]
include: 'external/docker-keeper/gitlab-ci-template.yml'

# Uncomment if ever you chose a different subtree prefix
# variables:
#   KEEPER_SUBTREE: external/docker-keeper
\end{lstlisting}
\end{boxedminipage}
\caption[Contents of docker-keeper-template/.gitlab-ci.yml]{Contents of \texttt{docker-keeper-template/.gitlab-ci.yml}.}
\label{fig:docker-keeper-template-ci-yml}
\end{figure}

The \texttt{images.yml} file gathers the specification of all Docker
images and tags. This specification is expressed in a domain-specific
language that I devised with a focus on genericity (albeit it is only
used for Coq projects to date).
This DSL is serialized in YAML, which
offers several upsides, in particular, it is a common file format in
the CI/CD ecosystem, it is more human-readable than JSON, and it supports
so-called anchors, aliases, and map merging features, which are handy
to ``factor-out'' common data and avoid maps duplicates in YAML
files.\footnote{cf.~%
  \url{https://docs.gitlab.com/ee/ci/yaml/yaml_optimization.html}
  for more details.}

\begin{landscape}
\begin{figure}
  \centering\frame{\includegraphics[width=1.72\textwidth]{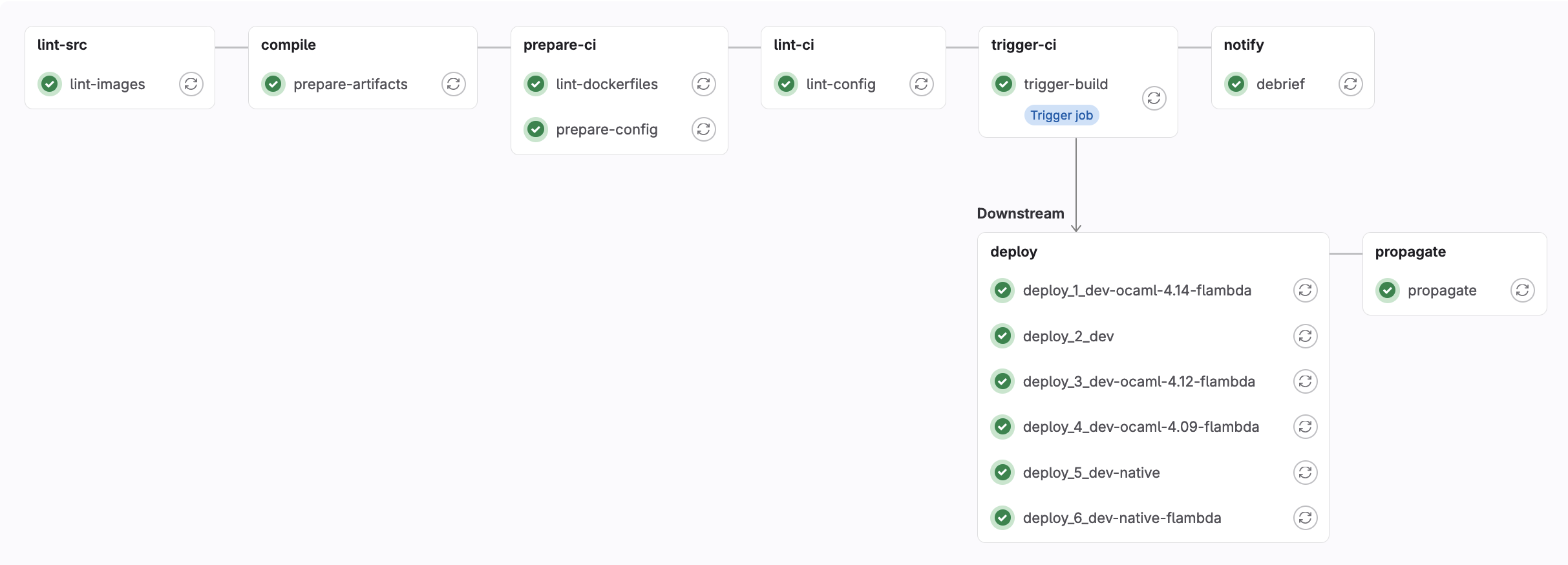}}
  \caption[Example pipeline (created by docker-keeper
    version 0.10.13)]{Example pipeline (created by docker-keeper
    \textsf{v0.10.13}): %
    the first row, which contains 6 stages.  The 5$^{\text{th}}$ stage
    triggers a so-called child pipeline (the 2$^{\text{nd}}$ row),
    which contains 2 stages. Each stage contains jobs that run in
    parallel. Stages run sequentially, left-to-right. A job failure
    prevents running the subsequent stages, and sets an error status
    for the pipeline. Jobs can be restarted, e.g., if they failed
    because of a temporary network error.}
  \label{fig:docker-keeper-pipeline}
\end{figure}
\end{landscape}

To give an overview of the DSL at stake,
Figure~\ref{fig:docker-mathcomp-images-yml} gives the beginning of the
\texttt{images.yml} file for the \texttt{docker-mathcomp} repository
(among the 3 repositories that rely on
docker-keeper --~cf.~Table~\ref{tab:docker-mathcomp}~--,
this is the simplest configuration).
A ``\texttt{keywords}'' based
labeling is available to be able to refer to several images in one
go, e.g., to easily rebuild all the images that are related to Coq
\texttt{8.19}. To be more precise, it would suffice to run the
following command to do so: \texttt{git commit -{}-allow-empty -m
  "docker-keeper:~rebuild-keyword:~8.19";~git push}. \\
Once docker-keeper runs, it produces a GitLab CI pipeline similar to
Figure~\ref{fig:docker-keeper-pipeline}.
Note that it is also possible to run the main steps of
docker-keeper locally, e.g., to double-check the order of tags in the
generated README.md file. To do so, we can run the commands described
in Figure~\ref{fig:docker-keeper-locally}.

Finally, it can be noted that some technical documentation for
docker-keeper is available in its
wiki.\footnote{cf.~\url{https://gitlab.com/erikmd/docker-keeper/-/wikis/home}
for docker-keeper's doc.}

\begin{figure}[!htb]
  \small
\begin{boxedminipage}{\textwidth}
\begin{verbatim}
# installation on first run
python3 -m venv "$PWD/venv"
. venv/bin/activate
python3 -m pip install -r external/docker-keeper/requirements.txt

# run docker-keeper locally
. venv/bin/activate
external/docker-keeper/keeper.py write-artifacts
cat generated/README.md # we advise adding generated/ in the .gitignore
\end{verbatim}
\end{boxedminipage}
\caption[Shell commands to run docker-keeper's main step
  locally]{Shell commands to run docker-keeper's main step
  locally, assuming we ``\texttt{cd}'' into the local Git repository
  that has docker-keeper as Git-subtree.}
\label{fig:docker-keeper-locally}
\end{figure}

\begin{figure}[!htb]
  \small
\begin{boxedminipage}{\textwidth}
\begin{lstlisting}[language=yaml]
---
base_url: 'https://gitlab.com/math-comp/docker-mathcomp'
active: true
gitlab_ci_tags:
  - 'large'
args:
  COQ_TAG: '{matrix[coq]}'
  MATHCOMP_VERSION: '{matrix[mathcomp]}'
docker_repo: 'mathcomp/mathcomp'
images:
  - matrix:
      mathcomp: ['2.2.0']
      coq: ['dev', '8.19', '8.18', '8.17', '8.16']
    build:
      keywords:
        - '{matrix[coq]}'
      context: './mathcomp'
      dockerfile: './single/Dockerfile'
      tags:
        - tag: '{matrix[mathcomp]}-coq-{matrix[coq]}'
        # TODO after next release: Remove "latest-coq-*"
        - tag: 'latest-coq-{matrix[coq]}'
  - matrix:
      mathcomp: ['2.1.0']
      coq: ['8.18', '8.17', '8.16']
    build:
      keywords:
        - '{matrix[coq]}'
      context: './mathcomp'
      dockerfile: './single/Dockerfile'
      tags:
        - tag: '{matrix[mathcomp]}-coq-{matrix[coq]}'
  # (...)
\end{lstlisting}
\end{boxedminipage}
\caption[Excerpt of
  docker-mathcomp/images.yml]{Excerpt of
  \texttt{docker-mathcomp/images.yml}~(cf.~\cite{docker-mathcomp}):
  the mandatory fields are \texttt{active}, \texttt{base\_url},
  \texttt{docker\_repo} (self-explanatory), and \texttt{images}, which
  is a sequence of YAML maps involving a \texttt{matrix} and a
  \texttt{build} specification. This latter map mentions at least
  the build \texttt{context} path, the source \texttt{dockerfile}, its
  build \texttt{args} (which are a standard feature to pass values
  from the host to the Dockerfile) here defined in a global way, and a
  list of \texttt{tags}. Many fields support some interpolation, e.g.,
  \YML{'\{matrix[coq]\}'}.}
\label{fig:docker-mathcomp-images-yml}
\end{figure}

\section{docker-keeper: inner workings}
\label{sec:docker-keeper-inner}

\begin{figure}[!htb]
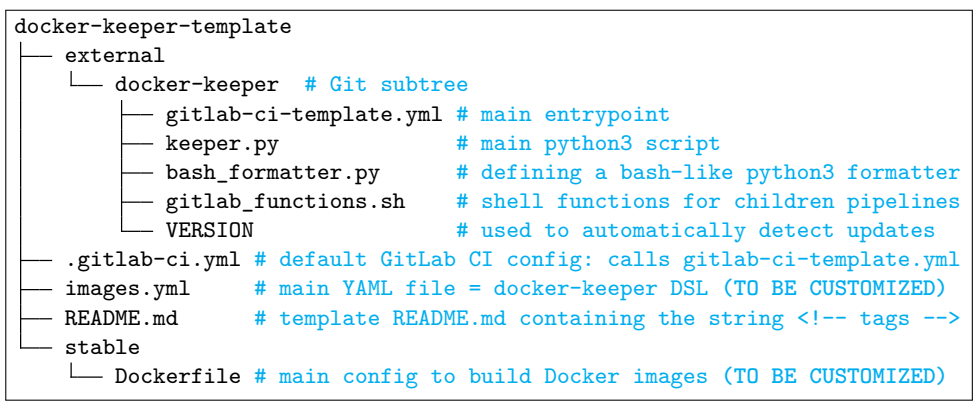

  \small
\!\!\!\begin{boxedminipage}{1.05\textwidth}
\begin{alltt}
docker-keeper-template
├── external
│   └── docker-keeper  {\color{cyan}# Git subtree}
│       ├── gitlab-ci-template.yml {\color{cyan}# main entrypoint}
│       ├── keeper.py              {\color{cyan}# main python3 script}
│       ├── bash_formatter.py      {\color{cyan}# defining a bash-like python3 formatter}
│       ├── gitlab_functions.sh    {\color{cyan}# shell functions for children pipelines}
│       └── VERSION                {\color{cyan}# used to automatically detect updates}
├── .gitlab-ci.yml {\color{cyan}# default GitLab CI config: calls gitlab-ci-template.yml}
├── images.yml     {\color{cyan}# main YAML file = docker-keeper DSL \textbf{(TO BE CUSTOMIZED)}}
├── README.md      {\color{cyan}# template README.md containing the string <!-- tags -->}
└── stable
    └── Dockerfile {\color{cyan}# main config to build Docker images \textbf{(TO BE CUSTOMIZED)}}
\end{alltt}
\end{boxedminipage}
\caption[Overview (continued) of the main files of
  docker-keeper-template]{Overview (continued) of the main files of the
  \texttt{docker-keeper-template} Git repository
  \cite{docker-keeper-template}.}
\label{fig:docker-keeper-template-subtree}
\end{figure}

Figure~\ref{fig:docker-keeper-template-subtree} shows the contents of
the \texttt{docker-keeper-template} repository (non-important files
are omitted) \cite{docker-keeper-template}. The technical architecture
of docker-keeper can be briefly summarized as follows. The main code
item is a Python 3 script (\texttt{keeper.py}) relying on the
\texttt{PyYAML} library, that generates GitLab CI children pipelines
(with one different job per Docker image to build). This script is
intended to be called by the main pipeline, but the question is, how
to easily provide the \texttt{keeper.py} script without cluttering the
main repository? We chose to rely on a Git subtree (more flexible that
Git submodules) and provide the maintainers with an optional scheduled
pipeline that checks whether the subtree should be updated w.r.t.\ the
upstream version.

\begin{figure}[!htb]
  \centering
  \includegraphics[width=0.8\textwidth]{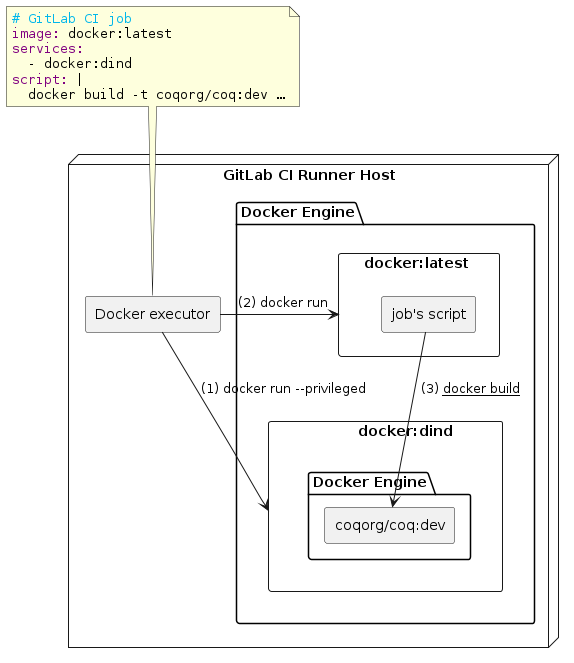}
  \caption[Architecture of GitLab CI jobs in a docker-keeper
    pipeline]{Architecture of GitLab CI jobs in a docker-keeper child
    pipeline: the Docker executor first runs a privileged container
    based on the Docker-in-Docker image (DinD) which thus comes with a
    nested Docker Engine, then runs another container based on the
    specified image, which finally runs the job's script that builds yet
    another image, cf. the underlined \underline{docker build} command
    in this figure.}
\label{fig:docker-keeper-dind}
\end{figure}

Next, let us review the main steps that occur within a docker-keeper pipeline.
Once such a pipeline is triggered (see
Figure~\ref{fig:docker-keeper-pipeline}), GitLab CI successively runs
the following stages:
\begin{itemize}
\item \textbf{lint-src}: lint \texttt{images.yml} using \texttt{yamllint};
\item \textbf{compile}: run \texttt{keeper.py write-artifacts \textrm{(with arguments\dots)}}
  to parse \texttt{images.yml}, compute the list of images to build
  with their synonymous tags, check that these tag names are disjoint,
  and store this data as an artifact; compute the list of referenced
  \texttt{Dockerfile}s and store this datum as an artifact; generate a
  comprehensive documentation from the repository \texttt{README.md}
  by replacing the pattern ``\texttt{<!-- tags -->}'' with the list of
  tags and links to their respective \texttt{Dockerfile};
\item \textbf{prepare-ci}: lint the \texttt{Dockerfile}s using
  \texttt{hadolint}; and in parallel, run \texttt{keeper.py
    generate-config} to generate a GitLab CI file \texttt{build.yml}
  gathering one separate job for each image to build, plus one special
  job \texttt{propagate} to trigger rebuilds in children docker-keeper
  repositories (if applicable);
\item \textbf{lint-ci}: lint the previously generated file
  \texttt{build.yml} using \texttt{yamllint};
\item \textbf{trigger-ci}: trigger a GitLab CI child pipeline from
  \texttt{build.yml};
\item \textbf{notify}: notify the maintainers with the list of old
  tags to remove from Docker Hub (if applicable), and the Markdown
  documentation to upload (which only changes when tags are added/removed).
\end{itemize}
It can be noted that in a docker-keeper child pipeline, each job runs
in a Docker container, and also has to run Docker commands for
building the target images: in GitLab CI, this is typically done using
Docker-in-Docker (DinD), see
Figure~\ref{fig:docker-keeper-dind}.

Now, let us elaborate a bit on the \texttt{propagate} feature. As
mentioned in Table~\ref{tab:docker-mathcomp}, the various
docker-keeper repositories depend on each other, so for example when a
docker-coq image is rebuilt, the related docker-mathcomp images should
be rebuilt downstream. For any GitLab CI instance, one can trigger new
pipelines by crafting a specific HTTP
request.\footnote{e.g.~\texttt{curl -X POST -F token="\$TOKEN" -F
    ref=master -F "variables[\$KEY1]=\$VALUE1"
    "https://\$GITLAB/api/v4/projects/\$ID/trigger/pipeline"}} The
methodology we followed consists in abstracting these requests,
extending the \texttt{images.yml} DSL with a declarative specification
of the requests, and the conditions when they are triggered. We
devised this language with the docker-coq/docker-mathcomp use cases in
mind, and with a focus on simplicity and conciseness.
Figure~\ref{fig:docker-coq-propagate} gives a comprehensive example of
a \texttt{propagate} instance (for docker-coq). It can be noted that
some YAML strings use non-standard format fields: %
\YML{'\{matrix[coq][//pl/.][\%.*]\}'}. We implemented this support of
Bash-like\footnote{cf. \texttt{\$\{VAR//pl/.\}} and
  \texttt{\$\{VAR\%.*\}} in Bash.} format fields in Python
so that we can very concisely
mean in this example: take the \texttt{matrix} variable, take the
value for its \texttt{coq} subfield (\texttt{matrix["coq"]} getting,
say, \texttt{"8.4pl6"}), replace all \texttt{"pl"} occurrences with
\texttt{"."}  (getting \texttt{"8.4.6"}), and drop the smallest 
suffix \texttt{".*"} (getting \texttt{"8.4"}).

\begin{figure}[!hb]
\!\!\!\begin{boxedminipage}{1.05\textwidth}
\small
\begin{lstlisting}[language=yaml]
docker_repo: 'coqorg/coq'
propagate:
  mathcomp:
    api_token_env_var: 'DMC_TOKEN'
    gitlab_domain: 'gitlab.inria.fr'
    gitlab_project: '44938'
    strategy:
      - when: 'rebuild-all'
        mode: 'rebuild-all'
      - when: 'forall'
        expr: '{matrix[coq][//pl/.][%.*]}'
        subset: '8.4,8.5'
        #§\color{cyan}$\iff\forall\textrm{b}\in\textrm{Builds}, \textrm{eval}(\texttt{"\{matrix[coq][//pl/.][\%.*]\}"}, \textrm{b})\subset\{\texttt{"}8.4\texttt{"},\texttt{"}8.5\texttt{"}\}$§
        mode: 'nil'
      - mode: 'rebuild-keyword'
        item: '{keywords[/#/,][#,]}'
        # this interpolation means: turn the list into a ,-separated string
  mathcomp-dev:
    api_token_env_var: 'MC_TOKEN'
    gitlab_domain: 'gitlab.inria.fr'
    gitlab_project: '44939'
    strategy:
      - when: 'rebuild-all'
        mode: 'minimal'
      - when: 'forall'
        expr: '{matrix[coq]}'
        subset: 'dev'
        mode: 'nightly'
      - when: 'exists'
        expr: '{matrix[coq][//pl/.][%.*]}'
        subset: '8.18,8.19,8.20,dev'
        #§\color{cyan}$\iff\exists\textrm{b}\in\textrm{Builds}, \textrm{eval}(\texttt{"\{matrix[coq][//pl/.][\%.*]\}"}, \textrm{b})\subset\{\textrm{\texttt{"}8.18\texttt{"},...,\texttt{"}dev\texttt{"}}\}$§
        mode: 'minimal'
\end{lstlisting}
\end{boxedminipage}
\caption[Excerpt of
  docker-coq/images.yml and propagate strategy]{Excerpt of
  \texttt{docker-coq/images.yml}~(cf.~\cite{docker-coq}). A
  \texttt{propagate strategy} defines a list of rules,
  with a
  condition \texttt{when:} (optional for the last rule) and a mandatory
  output \texttt{mode:} (to be applied to the triggered pipeline in
  the child repository). Conditions can be \texttt{rebuild-all} or
  \texttt{nightly} (input mode), or expressing a first-order
  condition of the form
  $\forall \textrm{build}.~E_1(\textrm{build}) \subset
  E_2(\textrm{build})$, or
  $\exists \textrm{build}.~E_1(\textrm{build}) \subset
  E_2(\textrm{build})$ where $E_1$ and $E_2$ are string expressions
  with interpolation, representing comma-separated lists. There is no
  explicit negation, but some implicit one: the order of rules matters, and
  the first condition that holds triggers the associated rule. Output
  modes can be \texttt{rebuild-all}, \texttt{rebuild-keyword},
  \texttt{nightly}, \texttt{minimal}, or  \texttt{nil} (i.e., don't propagate). }
\label{fig:docker-coq-propagate}
\end{figure}

\section{docker-coq-action: objectives and interface}
\label{sec:docker-coq-action-overview}

Up to 2019, CI/CD tasks for GitHub projects were generally performed
by using third-party services such as Travis CI or CircleCI. In late
2019, GitHub made its GitHub Actions service generally available,
which relies on YAML configurations called workflows. A workflow
defines a set of jobs executed on an Ubuntu, macOS, or Windows runner,
and a job itself consists in a sequence of steps, which are either a
shell command (field ``\texttt{run}'') or a so-called action (field
``\texttt{uses}''). Regarding the implementation of these actions,
there are two main kinds of actions: JavaScript-based actions (which
are multi-platform) and Container-based actions (which require a
GNU/Linux runner).

Given that most open-source Coq projects are hosted on GitHub
(including the projects of the Coq-community
organization\footnote{cf.~\url{https://coq-community.org/}}) and that
using GitHub's first-party CI/CD service looked like the way to go for
all such projects, the incentive to devise a proper GitHub-Actions
template for Coq projects was clear. Théo Zimmermann and I
collaboratively developed a new Container-based GitHub action, called
docker-coq-action~\cite{docker-coq-action}, based on the idea to
leverage the existing docker-coq images within GitHub workflows.

The implementation of docker-coq-action was non-trivial (I will
summarize its inner workings in the upcoming
Section~\ref{sec:docker-coq-action-inner}), but the outcome for end
users has clear upsides w.r.t.~the previous Travis~CI
counterpart:\footnote{cf.~\url{https://github.com/erikmd/docker-coq-travis-ci-demo-1/blob/master/.travis.yml}}
the user syntax is concise and intuitive, and it is expressive
enough to enable users to customize the build script either as a whole
(overriding ``\texttt{custom\_script}'') or part by part (only overriding
values among \G{before_install}, \mbox{\G{install},}
\mbox{\C{after_install},} \G{before_script}, \G{script},
\G{after_script}, or \G{uninstall}).
Figure \ref{fig:gha-template} gives a minimal example of use of
docker-coq-action, for a project entitled ``\texttt{coq-foo}'' that
would be tested against the latest release, and the development
version of Coq. Preparing artifacts for uploading them in a subsequent
step of the workflow, or adding variables to GitHub Actions'
environment files (namely, using the idiom %
\G{>>"\$GITHUB_ENV"}) is supported. The list of
available options is documented in the GitHub README of
docker-coq-action~\cite{docker-coq-action}.
%
It should be noted that although this action
contains ``coq'' in its name, it is language agnostic: it can be used to test projects in any language with
prebuilt Docker images. To do so, it suffices to omit the
\texttt{opam\_file} field and include the \texttt{custom\_script}
field.\footnote{cf.\;\url{https://github.com/coq-community/docker-coq-action/blob/v1.5.0/.github/workflows/python-demo.yml},
  for example.}

\begin{figure}[!htb]
\!\!\!\begin{boxedminipage}{1.05\textwidth}
\small
\begin{lstlisting}[language=yaml]
name: Docker CI
on:
  push:
    branches:
      - master
  pull_request:
    branches:
      - '**'
jobs:
  build:
    runs-on: ubuntu-latest
    strategy:
      matrix:
        image:
          - 'coqorg/coq:latest' # last stable release
          - 'coqorg/coq:dev'    # last master version
      fail-fast: false
    steps:
      - uses: actions/checkout@v4
      - uses: coq-community/docker-coq-action@v1
        with:
          opam_file: 'coq-foo.opam'
          custom_image: ${{ matrix.image }}
\end{lstlisting}
\end{boxedminipage}
\caption[Minimal complete example of GitHub workflow for a Coq project]{Minimal complete example of GitHub workflow for a Coq project
  applying the template from
  \url{https://github.com/coq-community/templates}: the
  \texttt{custom\_image} field gets each Docker image to run
  for the CI tests, while the \texttt{opam\_file} field gets the path of
  the \OPAM{} file of the Coq or OCaml project under test.}
\label{fig:gha-template}
\end{figure}

\section{docker-coq-action: inner workings}
\label{sec:docker-coq-action-inner}

\begin{figure}[!htb]
  \centering
  \includegraphics[width=0.8\textwidth]{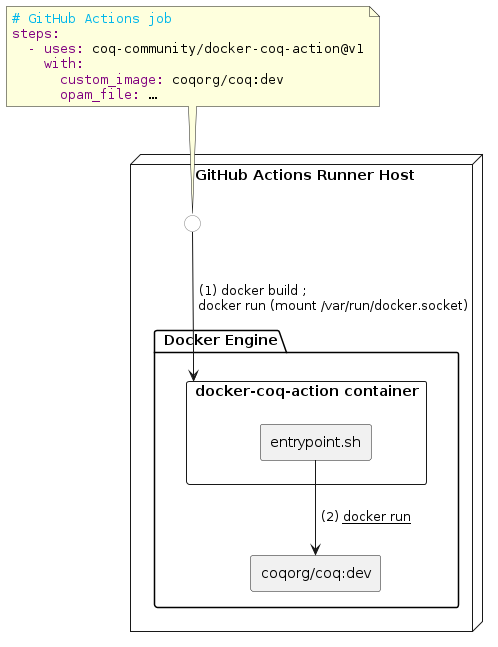}
  \caption[GitHub Actions: architecture of a Container-based
    action step]{GitHub Actions architecture of a step for a Container-based
    action: the runner successively builds the image associated with
    the Dockerfile and spins a container from this image with
    bind-mounting the host's Docker socket. Next, the underlying
    entrypoint of the container processes all the user-specified
    options, then runs yet another container from the specified
    \texttt{custom\_image}, cf.~the underlined \underline{docker run}
    command in this figure. In this Docker-out-of-Docker (DooD)
    approach, the last container is created in the host's Docker
    Engine, not inside a nested one.}
\label{fig:docker-coq-action-dood}
\end{figure}

As mentioned in the previous
Section~\ref{sec:docker-coq-action-overview}, the docker-coq-action is
a Container-based action: its main ingredients are thus a Dockerfile
(which is very simple, see~\ref{fig:docker-coq-action-dockerfile}) and
an entrypoint shell script. Roughly speaking, this action can be
viewed as an instance of the Facade design pattern for Docker-based CI
tests, with builtin support for \OPAM{} projects as well as for
several convenience features such as ``problem matchers'' (e.g., Coq
errors or warnings that appear in the log are recognized then
directly highlighted in the pull request diff view).
\begin{figure}[!htb]
\begin{boxedminipage}{\textwidth}
\begin{dockerfile}
FROM docker:latest # base image
WORKDIR /app       # §\textcolor{cyan}{$\iff$}§ mkdir -p /app; cd /app
COPY LICENSE README.md ./
COPY §entrypoint.sh§ timegroup.sh ./
COPY coq.json ./   # defines the regexp of Coq's errors/warnings
ENTRYPOINT ["/app/entrypoint.sh"]
\end{dockerfile}
\end{boxedminipage}
\caption[Dockerfile of the docker-coq-action]{Dockerfile of the docker-coq-action~\cite{docker-coq-action}.}
\label{fig:docker-coq-action-dockerfile}
\end{figure}
The overall behavior of a docker-coq-action step is summarized
in Figure~\ref{fig:docker-coq-action-dood}.

It can be noted that there is a common point with docker-keeper's
build jobs: the containerized script itself needs to run Docker
commands. Still, the runtime architecture of the two situations are
different: GitLab CI relies on the Docker-in-Docker approach (DinD,
see Figure~\ref{fig:docker-keeper-dind}), while the GitHub action
relies on the Docker-out-of-Docker approach (DooD, see
Figure~\ref{fig:docker-coq-action-dood}), reusing the Docker Engine of
the host.

Last, we would like to highlight the fact that the docker-coq-action
is being developed with a special focus on backward compatibility, so
that barring exceptional cases, new features should not lead to
breaking changes. The developing branch is named \texttt{"master"},
and we follow the standard convention that releases come with
\texttt{"v1.x.y"} tags, following {semanting versioning} (fixes just
increment \texttt{y}, the patch-level part of \texttt{v1.x.y}), and a
synonym reference \texttt{"v1"} directly points to the latest stable
release. But contrarily to some custom practice of GitHub Actions
maintainers, we do not change to which commit a tag points once it is
published. As a result, the latest stable version \texttt{"v1"} is not
implemented as a tag, but as a release branch.

\section{docker-coq contribs: related works, evaluation}
\label{sec:docker-coq-comparison}

Roughly speaking, our approach can be summarized with three main
characteristics: the use of containers, the use of \OPAM{} as package
management system, and the focus on the Coq proof assistant (albeit
docker-coq-action and docker-keeper are much more general).

In the sequel of this section, we highlight related works that may
apply for the same use case, and present a detailed comparison to
evaluate the strengths and the weaknesses of our approach. Next, we
will briefly review the verification effort performed for our
contributions, given a CI/CD tooling is itself a piece of software,
which can just as well be tested, and so on. Finally, we will suggest
perspectives for future work.

First, Théo Zimmermann developed a CI infrastructure for Coq projects,
based on the Nix package manager~\cite[Chap.~7]{Zimmermann:PhD}.
Then, some Nix support for the Mathematical Components library was developed by
Cyril
Cohen.\footnote{cf.~\url{https://github.com/math-comp/math-comp/wiki/Using-nix}}
Zimmermann's and Cohen's work led to the development and maintenance
of the
coq-nix-toolbox.\footnote{cf.~\url{https://github.com/coq-community/coq-nix-toolbox}}

Focusing more generally on the OCaml ecosystem (and still targeting
the GitHub CI/CD), a widely used way to setup a CI workflow amounts to
using
ocaml/setup-ocaml,\footnote{cf.~\url{https://github.com/ocaml/setup-ocaml}}
along with regular shell commands involving the \OPAM{} tool.  Given
the availability of the \OPAM{} tool in this approach, it is naturally
applicable for Coq projects.

In order to compare docker-coq-action/docker-coq with coq-nix-toolbox
and setup-ocaml, we first need to identify a relevant set of criteria.
Some previous works in software engineering have focused on comparing
CI/CD platforms (e.g., Jenkins vs.\ GitLab, by Singh et al.~\cite{Confluence2019}, or
Jenkins vs.\ TeamCity %
vs.\ Travis~CI, by Ivanov et al.~\cite{TEMJournal2022}). However, our approach is
different as we especially focus on one specific project language (the
Coq proof assistant) and on CI solutions compatible with the GitHub
platform (widely used for open-source projects).

We applied the following method: we enumerated salient characteristics
of each approach under study (in an unstructured manner), then
categorized them in general classes containing several different
criteria (ideally, two), which we finally use to evaluate the approaches in
a systematic way. These \textbf{classes} of \underline{\vphantom{q}criteria} are as follows:
\begin{description}
\item[applicability:] regarding the supported \underline{operating
    systems} and the number of \underline{available Coq/OCaml
    versions};
\item[generality:] regarding the availability of
  \underline{Coq-specific defaults} to apply on a Coq project
  out-of-the-box, and the \underline{genericity} to support languages
  beyond Coq;
\item[(Git/CI) coupling:] regarding the use of a \underline{single Git
    repository} (namely, a GitHub repository being the single source
  of truth) and the degree of \underline{integration of CI results} in
  the GitHub platform;
\item[(local/rev.) dependencies automation:] regarding the
  \underline{local reproducibility} of CI builds (whether it is
  feasible, or even fully automated with the dependencies prebuilt),
  and the facility to \underline{test and fix reverse dependencies}
  from the ecosystem (whether one such impact analysis is supported,
  and with what degree of automation);
\item[developer experience:] regarding the \underline{learning curve
    and ``skills lock-in''}\footnote{cf.~Hohpe~\cite{Hohpe2019}.}  and
  the \underline{readability} of CI scripts;
\item[performance:] regarding the maximum \underline{number of
    concurrent CI jobs} in a usual setting (platform-specific), and the
  \underline{elapsed time} of a typical CI job;
\item[reliability:] regarding the absence of ``false positives'' and
  ``false negatives'' for the CI pipeline status: this criterion is
  naturally important, but always achieved in practice (except in some
  very rare situations\footnote{see
    e.g.~\url{https://github.com/coq-community/coq-nix-toolbox/pull/162},
    where the combined use of two different CI approaches was useful
    to spot the issue.}), so we will omit it in the sequel.
\end{description}
It can be noted that criteria from the last three classes are close to
usual ones for comparing Turing-complete programming languages (see
e.g.~\cite[Chap.~1]{Claret:PhD}: ``programming languages can differ in
terms of readability, safety or efficiency.'')

\begin{figure}[!htb]
  \centering
  \includegraphics[width=\textwidth]{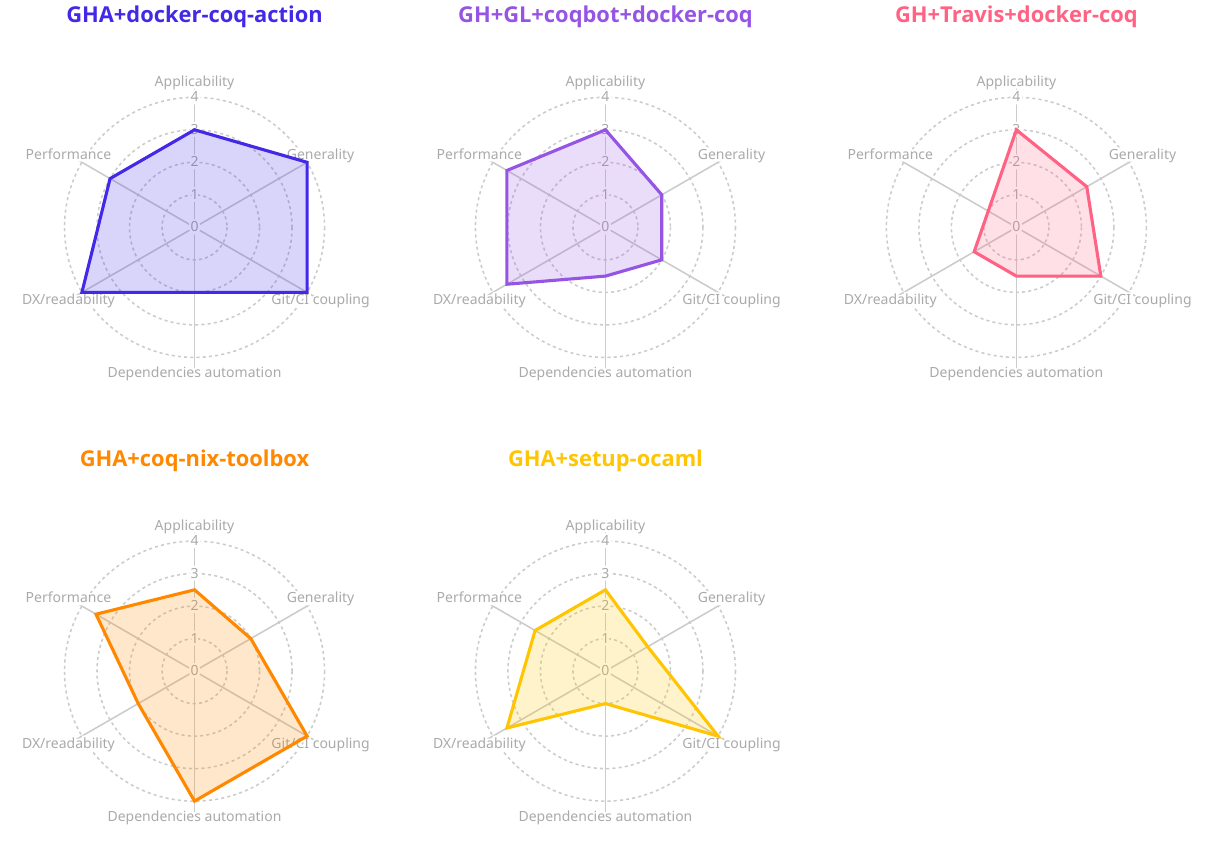}
  \caption[Radar char evaluating the 5 approaches w.r.t.\ our 6 criteria classes]{Radar char evaluating the 5 approaches w.r.t.\ our 6 criteria classes.}
\label{fig:docker-coq-radar}
\end{figure}

Figure~\ref{fig:docker-coq-radar} provides a visual overview of how
the 5 related approaches compare. (This radar chart is directly built
upon the data of Table~\ref{table:docker-coq-comparison-scale}, the
numerical scale of which being deduced from the comprehensive
comparison of the approaches w.r.t.\ our 12
\underline{\vphantom{q}criteria} presented in
Table~\ref{table:docker-coq-comparison}.)
We can see that docker-coq-action and coq-nix-toolbox cover a larger
area in the metrics space. Both approaches enjoy an optimal Git/CI
coupling score (they are naturally well-integrated in GitHub given
they are based on GitHub Actions). The coq-nix-toolbox approach excels
performance-wise as well as for automating the management of
local/reverse dependencies building and patching, while
docker-coq-action better performs w.r.t.\ the readability and
generality aspects (not only does docker-coq-action provide
Coq-specific defaults, but it is also applicable to any programming
language out-of-the-box). Also, docker-coq provides a binary cache for
slightly more combinations of Coq and OCaml versions.

\newif\ifscoreortext
\scoreortextfalse
\newcommand{\scoretext}[2]{\ifscoreortext #1\else #2\fi}
\newcommand{\gnulinux}{\scoretext{1}{use~a gnu-linux runner ($\neg$macos, $\neg$windows)}}
\newcommand{\linuxwin}{\scoretext{1.5}{use gnu-linux or macos runners (on windows: ``opam install coq'' fails)}}
\newcommand{\dockercoqocaml}{\scoretext{2}{($8.4\leq\text{Coq}\leq\text{dev}$)
    $\times$\footnote{cf.~\url{https://github.com/coq-community/docker-coq/wiki\#ocaml-versions-policy}}
    (upto 4 OCaml versions)}}
\newcommand{\dockercoqocamlbis}{\scoretext{2}{($8.4\leq\text{Coq}\leq\text{dev}$) $\times$ (upto 4 OCaml versions)}}
\newcommand{\nixcoqocaml}{\scoretext{1.5}{($8.5\leq\text{Coq}\leq\text{dev}$)
    $\times$\footnote{cf.\;\url{https://github.com/NixOS/nixpkgs/blob/24.05/pkgs/applications/science/logic/coq/default.nix\#L77-L86}}
    (1~default OCaml vers.)}}
\newcommand{\setupocamlcoq}{\scoretext{1}{any Coq/OCaml version but no central bin.~cache}}
\newcommand{\dockercoqactioniii}{\scoretext{2}{test-build a Coq/OCaml project out-of-the-box if an \texttt{.opam} file is provided}}
\newcommand{\dockercoqiii}{\scoretext{0.5}{a handcrafted CI script has to be provided}}
\newcommand{\setupocamliii}{\scoretext{0.5}{a handcrafted CI script has to be provided}}
\newcommand{\nixcoqiii}{\scoretext{1.5}{test-build a Coq project
    out-of-the-box if it has a \texttt{\_CoqProject}+\texttt{Makefile}
    and no extra dependency}}
\newcommand{\dockercoqactioniv}{\scoretext{2}{any prog.\ language with Docker images}}
\newcommand{\dockercoqtravisiv}{\scoretext{2}{any prog.\ language with Docker images}}
\newcommand{\dockercoqiv}{\scoretext{1.5}{any prog.\ language with Docker images,
    but GitLab sync lag drawback to be fixed by coqbot\footnote{cf.~\url{https://github.com/coq/bot/issues/234}}}}
\newcommand{\nixcoqiv}{\scoretext{0.5}{the implemented infra\-structure is Coq specific albeit the approach could be generalized}}
\newcommand{\setupocamliv}{\scoretext{1}{any project packaged with \OPAM}}
\newcommand{\dockercoqactionv}{\scoretext{2}{1 self-contained repo.}}
\newcommand{\dockercoqv}{\scoretext{1}{need 2 repos per project (GitHub+GitLab)}}
\newcommand{\dockercoqtravisv}{\scoretext{2}{1 self-contained repo.}}
\newcommand{\nixcoqv}{\scoretext{2}{1 self-contained repo.}}
\newcommand{\setupocamlv}{\scoretext{2}{1 self-contained repo.}}
\newcommand{\dockercoqactionvi}{\scoretext{2}{well-integrated: pull request problem matchers, status checks}}
\newcommand{\dockercoqvi}{\scoretext{1}{integrated with GitHub's status checks}}
\newcommand{\dockercoqtravisvi}{\scoretext{1}{integrated with GitHub's status checks}}
\newcommand{\nixcoqvi}{\scoretext{2}{well-integrated: pull request problem matchers, status checks}}
\newcommand{\setupocamlvi}{\scoretext{2}{well-integrated: pull request problem matchers, status checks}}
\newcommand{\dockercoqactionvii}{\scoretext{1}{minimal (the base Docker image can be pulled)}}
\newcommand{\dockercoqvii}{\scoretext{1}{minimal (the base Docker image can be pulled)}}
\newcommand{\dockercoqtravisvii}{\scoretext{1}{minimal (the base Docker image can be pulled)}}
\newcommand{\nixcoqvii}{\scoretext{2}{optimal (Nix can reuse artifacts cached in CI)}}
\newcommand{\setupocamlvii}{\scoretext{0.5}{possible in theory, but no support}}
\newcommand{\dockercoqactionviii}{\scoretext{1}{some support in coq-community templates\footnote{cf.~\url{https://github.com/coq-community/templates/pull/118}}}}
\newcommand{\dockercoqviii}{\scoretext{0.5}{possible in theory, but no support}}
\newcommand{\dockercoqtravisviii}{\scoretext{0.5}{possible in theory, but no support}}
\newcommand{\nixcoqviii}{\scoretext{2}{support for reverse dependencies testing and so-called overlays\footnote{cf.~\url{https://github.com/coq-community/coq-nix-toolbox\#overlays}}}}
\newcommand{\setupocamlviii}{\scoretext{0.5}{possible in theory, but no support}}
\newcommand{\dockercoqactionix}{\scoretext{2}{very concise and readable}}
\newcommand{\dockercoqix}{\scoretext{1.5}{concise and readable}}
\newcommand{\dockercoqtravisix}{\scoretext{0.5}{verbose (manual) code}}
\newcommand{\nixcoqix}{\scoretext{1}{verbose (generated) code}}
\newcommand{\setupocamlix}{\scoretext{1.5}{concise and readable}}
\newcommand{\dockercoqactionx}{\scoretext{2}{Bash knowledge suffices to customize build}}
\newcommand{\dockercoqx}{\scoretext{2}{Bash knowledge suffices to customize build}}
\newcommand{\dockercoqtravisx}{\scoretext{1}{need Docker+Bash skills}}
\newcommand{\nixcoqx}{\scoretext{1}{need familiarity with Nix}}
\newcommand{\setupocamlx}{\scoretext{2}{Bash knowledge suffices to customize build}}
\newcommand{\dockercoqactionxi}{\scoretext{1.5}{20 jobs (in the Free plan)}}
\newcommand{\dockercoqxi}{\scoretext{2}{500 jobs on gitlab.com\footnote{cf.\url{https://docs.gitlab.com/administration/instance_limits/}},
    can be manually reduced}} %
\newcommand{\dockercoqtravisxi}{\scoretext{0.5}{0 (no free tier anymore)}}
\newcommand{\nixcoqxi}{\scoretext{1.5}{20 jobs (in the Free plan)}}
\newcommand{\setupocamlxi}{\scoretext{1.5}{20 jobs (in the Free plan)}}
\newcommand{\dockercoqactionxii}{\scoretext{1.5}{6' (3' with a bigger image)}}
\newcommand{\dockercoqxii}{\scoretext{1.5}{should be similar to GHA (untested)}}
\newcommand{\dockercoqtravisxii}{\scoretext{0.5}{N/A (coudn't reproduce)}}
\newcommand{\nixcoqxii}{\scoretext{2}{1' (using Cachix)}}
\newcommand{\setupocamlxii}{\scoretext{1}{18' (12' once cache hits)}}

\newcommand{\TableGHARadar}[2][]{\begin{landscapetable}
\begin{scriptsize}
\begin{tabular}{p{.15\linewidth}p{.15\linewidth}p{.15\linewidth}p{.15\linewidth}|p{.15\linewidth}p{.15\linewidth}p{.15\linewidth}}
  \toprule
  Approach: & GitHub\,Actions+\linebreak docker-coq-action & GitHub+GitLab+\linebreak coqbot+docker-coq & GitHub+Travis\,CI+\linebreak docker-coq & GitHub\,Actions+\linebreak coq-nix-toolbox & GitHub\,Actions+\linebreak setup-ocaml \\ \midrule
  #1
  1.~operating systems & \gnulinux & \gnulinux & \gnulinux & \gnulinux & \linuxwin \\
  2.~\# available versions of Coq and OCaml & \dockercoqocaml & \dockercoqocamlbis & \dockercoqocamlbis & \nixcoqocaml & \setupocamlcoq \\ \midrule
  3.~Coq-specific defaults & \dockercoqactioniii & \dockercoqiii & \dockercoqiii & \nixcoqiii & \setupocamliii \\
  4.~genericity & \dockercoqactioniv & \dockercoqiv & \dockercoqtravisiv & \nixcoqiv & \setupocamliv \\ \midrule
  5.~use of a single Git repository & \dockercoqactionv & \dockercoqv & \dockercoqtravisv & \nixcoqv & \setupocamlv \\
  6.~integration of CI results in GitHub & \dockercoqactionvi & \dockercoqvi & \dockercoqtravisvi & \nixcoqvi & \setupocamlvi \\ \midrule
  7.~local reproducibility automation & \dockercoqactionvii & \dockercoqvii & \dockercoqtravisvii & \nixcoqvii & \setupocamlvii \\
  8.~reverse-dependencies testing automation & \dockercoqactionviii & \dockercoqviii & \dockercoqtravisviii & \nixcoqviii & \setupocamlviii \\ \midrule
  9.~CI script readability 
    & \dockercoqactionix & \dockercoqix & \dockercoqtravisix & \nixcoqix & \setupocamlix \\
  10.~learning curve and skills lock-in & \dockercoqactionx & \dockercoqx & \dockercoqtravisx & \nixcoqx & \setupocamlx \\ \midrule
  11.~maximum~number of concurrent CI jobs & \dockercoqactionxi & \dockercoqxi & \dockercoqtravisxi & \nixcoqxi & \setupocamlxi \\
  12.~elapsed time of a typical CI job & \dockercoqactionxii & \dockercoqxii & \dockercoqtravisxii & \nixcoqxii & \setupocamlxii \\ \bottomrule
\end{tabular}
#2
\end{scriptsize}
\end{landscapetable}}

\scoreortexttrue
\TableGHARadar{\caption[Comparaison of the 5 approaches on a numerical
    scale]{Comparaison of the 5 approaches on a numerical
    scale, deduced from Table~\ref{table:docker-coq-comparison} and used for
    Figure~\ref{fig:docker-coq-radar}.}\label{table:docker-coq-comparison-scale}}
\scoreortextfalse
\TableGHARadar[%
Project example relying on this CI approach
  & UniMath\footnote{cf.~\url{https://github.com/UniMath/UniMath}}
  & MathComp\footnote{cf.~\url{https://gitlab.inria.fr/math-comp/math-comp}}
  & docker-coq-travis-ci-demo-1\,\footnote{cf.\;\url{https://github.com/erikmd/docker-coq-travis-ci-demo-1/}~---~%
    discontinued (no more free tier for open source projects in Travis CI since December 2020)}
  & coq-mathcomp-analysis\footnote{cf.~\url{https://github.com/math-comp/analysis}}
  & old version of UniMath\footnote{cf.\;the ubuntu/macos CI scripts before
    \url{https://github.com/UniMath/UniMath/pull/1599}} \\ \midrule
]{\caption[Detailed comparison of the 5 related approaches]{Detailed comparison of the 5 related approaches (as of
    September 2024).}\label{table:docker-coq-comparison}}

\section{Conclusion and perspectives}
\label{sec:conclusion}

Docker-Coq images are available since Nov.~2018 and have been
pulled $1.007\cdot10^6$ times in (slightly less than) 7 years, according to
the Docker Hub API.

The docker-coq-action is distributed in the GitHub Actions marketplace
since April 2020. It nicely complements the availability of these
images by providing a concise and human-friendly way to build and test
a Rocq project. It can either be directly used from a handcrafted
GitHub workflow, or as part of the Rocq-community
templates.\footnote{cf.~\url{https://github.com/rocq-community/templates},
  based on a \texttt{meta.yml} ``pivot file''.}  The docker-coq-action
is not specific to the Coq/Rocq ecosystem though, given vanilla OCaml
projects can rely on it just as well.  As of September 2025,
docker-coq-action is used by 501~public GitHub repositories, according
to the GitHub API.

The docker-keeper tooling is even more generic. It has been developed
since June 2020 to compensate for the lack of scalability and
maintainability offered by Docker Hub's automated
builds\footnote{cf.~\url{https://docs.docker.com/docker-hub/builds/}}
for a compiler project like Rocq. The docker-keeper project combines a
YAML user-friendly DSL, and a Python3 compiler endowed with a
Bash-like formatter to easily define rules for synonymous tags, as
well as
cross-project propagation. It has
been developed with a special focus on reliability, by adding several
linting passes to check the various input files (e.g., Dockerfiles) as well as
the generated YAMLs that subsequently trigger GitLab CI children pipelines. But if the
correctness of docker-keeper is deemed as critical, it would be
feasible to rewrite it into a strongly typed functional
programming language such as OCaml or Rocq, to increase type safety, or
even prove correctness properties (while keeping the same DSL to allow
for a drop-in replacement).

Regarding related works, docker-keeper is similar to %
bashbrew\footnote{cf.~\url{https://github.com/docker-library/bashbrew}},
which is the main building block for (re)building Docker's Official
Images\footnote{cf.~\url{https://hub.docker.com/u/library}}. The main
difference is that docker-keeper is a lightweight compiler that
specifically targets GitLab CI, while bashbrew targets Jenkins.
Another interesting related work is the ocurrent
project\footnote{cf.~\url{https://github.com/ocurrent/ocurrent}}
(developed since May 2019), which is currently used to maintain the
OCaml Docker base images and the CI of
opam-repository\footnote{cf.~\url{https://github.com/ocaml/opam-repository/}}.
It has more flexibility (build steps can be vanilly OCaml scripts),
but it is heavier than docker-keeper to setup and host.

Regarding the renaming from Coq to Rocq, a large part of the tweaks
required by the renaming have already been applied (e.g., creating
the GitHub repository
\href{https://github.com/rocq-community/docker-rocq}{\nolinkurl{rocq-community/docker-rocq}}
and the two Docker Hub repositories
\href{https://hub.docker.com/r/rocq/base}{\nolinkurl{rocq/base}} and
\href{https://hub.docker.com/r/rocq/rocq-prover}{\nolinkurl{rocq/rocq-prover}})
so that images for Coq $\leqslant 8.20.1$ belong to
\href{https://hub.docker.com/r/coqorg/coq}{\nolinkurl{coqorg/coq}}
($\Leftrightarrow$ docker-coq) and
images for Rocq $\geqslant 9.0.0$ belong to
\href{https://hub.docker.com/r/rocq/rocq-prover}{\nolinkurl{rocq/rocq-prover}}
($\Leftrightarrow$ docker-rocq).
Still, some migration steps remain (e.g., renaming the GitHub
repository
\href{https://github.com/rocq-community/docker-coq-action}{\nolinkurl{rocq-community/docker-coq-action}}
into \texttt{rocq-prover/docker-opam-action}, renaming its
\texttt{COQ\_IMAGE} environment variable into \texttt{DOCKER\_IMAGE}
while releasing \texttt{v2} to perform this non-backwards-compatible
change). Also, feature wishes are listed as docker-keeper issues.\footnote{cf.~\url{https://gitlab.com/erikmd/docker-keeper/-/issues}}

We hope this document will help docker-coq-action and docker-keeper to
be further extended and used.

\section*{Acknowledgments}
The author would like to thank the members of the Rocq dev team as
well as of the math-comp dev team for many useful discussions, and in
particular many thanks to Théo Zimmermann, Pierre Roux, Karl Palmskog
(the current owners of the Rocq-community project) for their advice
and help. Special thanks go to Jason Gross for his numerous and
relevant suggestions in the docker-coq-action issues tracker.\footnote{cf.~\url{https://github.com/rocq-community/docker-coq-action/issues}}

\registerbmtoc{List of figures}
\listoffigures

\registerbmtoc{List of tables}
\listoftables

\registerbmtoc{References}
\bibliographystyle{plainurl}
\bibliography{biblio}

\end{document}